\documentclass[%
11pt,
reprint,
onecolumn,
tightenlines,
superscriptaddress,
preprintnumbers,
nofootinbib,
amsmath,amssymb,amsthm,
physrev,
eqsecnum,tikz,
]{revtex4-2}

\usepackage{isomath}
\usepackage{amsmath,amsthm}
\usepackage{amsbsy}
\usepackage{amssymb}
\usepackage{amscd}
\usepackage{amsfonts}
\usepackage{stmaryrd}
\usepackage{siunitx}
\usepackage{euscript}
\usepackage[utf8]{inputenc}
\usepackage[T1]{fontenc}
\usepackage{newtxtext} 
\everymath{\displaystyle}
\usepackage{exscale}
\usepackage{microtype}
\usepackage{hyperref}
\usepackage{booktabs}
\usepackage{algorithm}
\usepackage{algpseudocode}
\usepackage{newtxmath} 
\usepackage[scr=rsfso]{mathalfa}
\usepackage{url}

\usepackage{graphicx}
\usepackage{boxedminipage}
\usepackage{calc}
\usepackage[usenames,dvipsnames]{xcolor}
\graphicspath{ {media/} }
\usepackage[caption=false,justification=centerlast]{subfig}
\usepackage{tabularx}
\usepackage{array,multirow,graphicx, makecell}
\usepackage{tikz}

\usepackage{setspace}
\usepackage{enumitem}
\setitemize{noitemsep,topsep=0pt,parsep=0pt,partopsep=0pt}
\setenumerate{noitemsep,topsep=0pt,parsep=0pt,partopsep=0pt}
\setdescription{noitemsep,topsep=0pt,parsep=0pt,partopsep=0pt}

\usepackage[colorinlistoftodos, color=green!40,prependcaption]{todonotes}
\setuptodonotes{inline}

\usepackage{soul} 
\usepackage[normalem]{ulem}

\usepackage{orcidlink}
\usepackage{siunitx}
\usepackage[small]{titlesec}

\titlespacing*{\section}{0pt}{12pt plus 4pt minus 2pt}{2pt plus 2pt minus 2pt}
\titlespacing*{\subsection}{0pt}{12pt plus 4pt minus 2pt}{2pt plus 2pt minus 2pt}
\titlespacing*\subsubsection{0pt}{12pt plus 4pt minus 2pt}{2pt plus 2pt minus 2pt}
\titlespacing*\paragraph{0pt}{12pt plus 4pt minus 2pt}{2pt plus 2pt minus 2pt}

\makeatletter
    \renewcommand*{\thesection}{\arabic{section}}
    \renewcommand*{\thesubsection}{\thesection.\Alph{subsection}}
    \renewcommand*{\p@subsection}{}
    \renewcommand*{\thesubsubsection}{\thesubsection.\arabic{subsubsection}}
    \renewcommand*{\p@subsubsection}{}
\makeatother

\usepackage{isomath}
\usepackage{amsmath}
\usepackage{amssymb}
\usepackage{amscd}
\usepackage{amsfonts}
\usepackage{upgreek}

\newcommand{\R}{\mathbb R}

\theoremstyle{definition}

\AtEndEnvironment{remark}{\null\hfill\qedsymbol}

\AtEndEnvironment{example}{\null\hfill\qedsymbol}

\newcommand{\bfxi}{\mathbold {\xi}}

\newcommand\myatop[2]{\genfrac{}{}{0pt}{}{#1\hfill}{#2\hfill}}

\newcommand{\bfg}{{\mathbold g}}

\newcommand{\bfn}{{\mathbold n}}

\newcommand{\bfx}{{\mathbold x}}

\newcommand{\bfP}{{\mathbold P}}
\newcommand{\bfQ}{{\mathbold Q}}

\begin{document}


\preprint{To appear in Physical Review Materials (\url{https://doi.org/10.1103/PhysRevMaterials.8.093403})}

\title{Impact of Grain Boundary Energy Anisotropy on Grain Growth}

\author{S. Kiana Naghibzadeh \orcidlink{0000-0001-5629-9396}}
    \email{kiana@mit.edu}
    \affiliation{Department of Civil and Environmental Engineering, Massachusetts Institute of Technology}

\author{Zipeng Xu}
    \affiliation{Department of Materials Science and Engineering, Carnegie Mellon University}

\author{David Kinderlehrer \orcidlink{0000-0001-7598-5762}}
    \affiliation{Center for Nonlinear Analysis, Department of Mathematical Sciences, Carnegie Mellon University}

\author{Robert Suter\orcidlink{0000-0002-0651-0437}}
    \affiliation{Department of Physics, Carnegie Mellon University}
    \affiliation{Department of Materials Science and Engineering, Carnegie Mellon University}

\author{Kaushik Dayal \orcidlink{0000-0002-0516-3066}}
    \affiliation{Department of Civil and Environmental Engineering, Carnegie Mellon University}
    \affiliation{Center for Nonlinear Analysis, Department of Mathematical Sciences, Carnegie Mellon University}
    \affiliation{Department of Mechanical Engineering, Carnegie Mellon University}

\author{Gregory S. Rohrer \orcidlink{0000-0002-9671-3034}}
    \affiliation{Department of Materials Science and Engineering, Carnegie Mellon University}

\date{\today}


\begin{abstract}
    A threshold dynamics model of grain growth that accounts for the anisotropy in the grain boundary energy has been used to simulate experimentally observed grain growth of polycrystalline Ni.  
    The simulation reproduces several aspects of the observed microstructural evolution that are not found in the results of simulations assuming isotropic properties.  
    For example, the relative areas of the lowest energy twin boundaries increase as the grains grow and the average grain boundary energy decreases with grain growth.  
    This decrease in energy occurs because 
    the population of higher energy grain boundaries
    decreases while the population of lower energy boundaries increases
    as the total grain boundary area decreases. 
    This phenomenon emerges from the assumption of anisotropic grain boundary energies without modification of the energy minimizing algorithm.
    These findings are consistent with the observation that in addition to the decrease in grain boundary area, additional energy is dissipated during grain growth by a decrease in the average grain boundary energy.

\end{abstract}

\maketitle


\section{Introduction}
Grain boundaries are the interfaces between crystals with different lattice orientations in polycrystalline metals, ceramics, polymers, and rocks. 
At high temperatures, grain boundaries migrate and this is one important mechanism for the evolution of polycrystalline microstructures. 
Grain growth, which is an increase in the average crystal size by grain boundary migration, affects structure-sensitive material properties.
Hence, understanding the underlying mechanism of grain boundary migration is necessary for controlling 
the electrical, optical, and mechanical properties of materials.
Grain growth by grain boundary migration has been extensively studied in the past using 
analytical theories \cite{hillert1965theory, fischer2008distribution, gottstein1998grain}, 
molecular dynamics simulations \cite{holm2010grain, jhan1990molecular, upmanyu1999triple, chen2020atomistic, bizana2023kinetics, race2014role, zhang2006characterization, chen2018atomistic}, Monte Carlo simulations \cite{srolovitz1984computer, srolovitz1986grain,  anderson1989computer}, phase field simulations \cite{chen2002phase, kobayashi1998vector, mckenna2014grain, moelans2022new, moelans2008quantitative,ito2019bayesian, kamachali20123, kazaryan2001grain, bjerre2013rotation, adland2013unified, admal2019three, krill2002computer, korbuly2017grain, tang2006diffuse, vanherpe2007bounding}, threshold dynamics \cite{peng2022comparison, nino2023influence} and other approaches \cite{mullins1993scaling, humphreys2000modelling, barmak2011critical, bardsley2017towards, bragg1947dynamical, farjas2007numerical, moldovan2002scaling, chen2017effect, gruber2005effect, florez2022statistical}.

Recent experimental observations have provided two findings not captured by most of the simulations.  
The first is that grain boundaries are approximately equally likely to migrate toward or away from their centers of curvature \cite{muralikrishnan2023observations, xu2024grain, bhattacharya2021grain}.
The second is that while grain boundaries move to decrease the total energy of the system by decreasing the grain boundary area, they further decrease the energy by replacing high-energy grain boundaries with low-energy ones, a process referred to as grain boundary replacement \cite{xu2023energy}. This suggests that grain boundary energy must be included in the simulations.
The grain boundary energy (GBE) depends on five macroscopic parameters, which can be expressed as the lattice misorientation between the two adjacent grains (three degrees of freedom) forming the boundary, and the orientation of the boundary plane (two degrees of freedom) \cite{rohrer2011grain, ratanaphan2015grain}. 
However, most previous grain growth simulations considered isotropic grain boundary energy, i.e. the energy is the same for all boundaries, and this cannot capture the replacement of high-energy grain boundaries with low-energy ones.

Most of the simulations that used anisotropic properties only considered the dependency of GBE on the misorientation 
\cite{elsey2013simulations, hallberg2014influence, esedoḡ2015threshold} or simulated grain growth in two dimensions \cite{florez2022statistical}.
However, the GBE varies more strongly with variations in the grain boundary plane orientation than with the lattice misorientation \cite{rohrer2011grain}, so it seems unlikely that simulations ignoring these parameters will correctly simulate the energy reduction during grain growth.  

Two prior examples of three-dimensional simulations considering the grain boundary plane dependence on the GBE in three-dimensions considered a hypothetical GBE function and did not include the dependence of the GBE on the lattice misorientation \cite{gruber2005effect, salama2020role}.
A limited number of recent studies have simulated grain growth in 3D using a GBE that varies with all five parameters \cite{kim2014phase, hallberg2019modeling, nino2023influence}.
Kim et al. \cite{kim2014phase} used the phase field method to simulate grain growth in BCC Fe and found that the anisotropic GBE influenced the morphological evolution of grains and that low energy boundaries increased in population during grain growth. 
Hallberg and Bulatov \cite{hallberg2019modeling} developed an anisotropic level set method to show the importance of energy anisotropy in the morphology of evolved microstructures. 
The simulations used energies specified by the Bulatov-Reed-Kumar (BRK) energy function \cite{bulatov2014grain} for fcc structures and models no more than four grains.
Nino and Johnson \cite{nino2023influence} used a simplified extension of threshold dynamics (TD) with different energy functions to study the effect of energy anisotropy on the evolution of a microstructure instantiated with a Voronoi tessellation. This simplified version employed Gaussian kernels to describe the GBE of boundaries of different inclinations. A discussion on the difference between the current method and an extension used by \cite{nino2023influence} is provided in section \ref{sec:TD}.


The purpose of this study was to compare the outcome of the simulated 3D microstructure evolution 
with the experimentally observed evolution of Ni during grain growth \cite{hefferan2010statistics, bhattacharya2019three} using both an isotropic and a five-parameter anisotropic GBE.  
The simulations described here differ from the previous work in two ways. 
The first is that an improved description of the grain boundary energy anisotropy, described in section \ref{sec:TD}, is implemented. 
The second is that experimentally observed microstructures are used as input, and the results are compared to the observed microstructure in later states.  
The experimentally determined starting microstructures contain 2400 to 3000 grains, and we quantify changes in the average grain boundary energy and the grain boundary energy distribution, the grain boundary area, the grain boundary curvature, the grain boundary velocity, changes in the numbers of near neighbors, and the relative areas of twin boundaries.  The simulation captures the replacement of higher energy GBs by lower energy boundaries in the same manner that was observed experimentally \cite{xu2023energy}, while the isotropic simulation cannot necessarily predict this mechanism.

To simulate grain growth, we have used the TD method originally introduced by Merriman,
Bence and Osher in \cite{merriman1992diffusion,merriman1994motion} which uses an implicit representation of boundaries. 
There are three reasons for choosing the TD method.  
The first is the high computational efficiency compared to other methods using an implicit representation of the interface, such as the phase field method and level set method.  
The second reason is the straightforward extension of the model to anisotropic simulations using experimentally derived interface properties. 
The third is that the data structure of the model is analogous to that of the experiment, allowing data to be easily transferred and the analysis of microstructural characteristics to be computed using the same codes for the experimental and simulated data. 
Furthermore, Nino and Johnson \cite{nino2023influence} showed that anisotropic threshold dynamics simulations produced triple junction geometries that were consistent with the Herring \cite{herring1999surface} condition.

There are different methods to incorporate the experimentally derived grain boundary energy into the evolution algorithm \cite{bonnetier2012consistency, elsey2018threshold, esedoḡlu2017kernels, ishii1999threshold, esedoglu2017convolution}. 
To consider a fully anisotropic TD method, we will follow \cite{bonnetier2012consistency}, because it has no restriction on the choice of the grain boundary energy function that can be considered, and it is computationally less expensive than other available models as it only requires the GBE itself and not its derivatives.
To evaluate the grain boundary energy, the five-parameter grain boundary function defined by Bulatov et al. (Bulatov-Reed-Kumar (BRK) energy function) is used \cite{bulatov2014grain}.
Although this function is only an interpolation between 388 calculated GBE values from molecular dynamics simulation \cite{olmsted2009survey}, it has been shown to be a good approximation of experimentally determined GBEs \cite{rohrer2010comparing}.

\section{Materials and Methods}

\subsection{Threshold dynamics}
\label{sec:TD}
The threshold dynamics (TD) algorithm is a method to simulate free boundary motion by mean curvature and was initially introduced by Merriman, Bence, and Osher in \cite{merriman1992diffusion, merriman1994motion}.
In this approach, each grain $i$ is identified by a characteristic function $\mathbf{1}_{\Sigma_{i}^{k}}$ which has the value one within grain $i$ and zero outside the grain. The set of position vectors within grain $i$ at time $t_k$  are denoted by $\Sigma_{i}^{k}$. 
To evolve a microstructure with $N$ grains at time $t_k$ and evaluate the microstructure at time $t_{k+1}>t_k$, Algorithm \ref{alg:TD} \cite{elsey2018threshold, esedoḡ2015threshold} is used.
In the convolution step, each grain’s characteristic function is convolved with a kernel specific to each boundary, $K_{\delta t}^{i,j}=K_{\delta t}^{j,i}$. 
Typically, the kernels are defined such that they are maximum at the origin and decay to zero at infinity, the overall rate of the decay to zero depends on $\delta t$ and the rate of the decay in each direction depends on the anisotropy of the grain boundary energy as described below \cite{merriman2000convolution}. 
At each point $\bfx$, the convolution operator computes the integral of the product between the characteristic function and the kernel, with the kernel re-positioned so that its maximum is located at point $\bfx$. 
Hence, the value of the convolution $\psi_i^k$ at point $\bfx$ deep inside grain $i$ and far from the boundary remains zero while its value increases as $\bfx$ gets closer to the boundary and increases further outside of the grain. Note that the convolution value $\psi_i^k$ highly depends on the curvature of the boundary as the convolution which results from the overlap between the non-zero part of the kernel and the non-zero part of the characteristic function $\mathbf{1}_{\Sigma_{j}^{k}}$, can vary significantly depending on the shape of the grain.  
In the thresholding step, the characteristic function for each grain is redefined such that $\mathbf{1}_{\Sigma_{i}^{k+1}}$ at time $t_{k+1}$  is equal to one at points where $\psi_i$ is minimum compared to $\psi_j$ for all other grains with $j=1,...,N$, and zero otherwise.
This way, the boundaries are moved effectively by weighted mean curvature.

\begin{algorithm}[H]
    \caption{Anisotropic Threshold dynamics}
    \label{alg:TD}
    \textbf{Initialization:}\ 
    Given $\Sigma_1^k, ... \Sigma_N^k$ and time step size $\delta t$
    \\
    \textbf{Convolution:}
    \begin{equation}
        \label{eqn:convolution}
        \psi_{i}^{k}=\sum_{\myatop{j=1}{j \neq i}}^{N} K_{\delta t}^{i, j} * \mathbf{1}_{\Sigma_{j}^{k}}
    \end{equation}
    \textbf{Thresholding:}
    \begin{equation}
        \label{eqn:thresholding}
        \Sigma_{i}^{k+1}=\left\{\bfx: \psi_{i}^{k}(\bfx) \leq \min _{j \neq i} \psi_{j}^{k}(\bfx)\right\}
    \end{equation}
\end{algorithm}

In the generalized form of the algorithm \cite{esedoḡ2015threshold}, $K_{\delta t}^{i,j}$ can be different for different grain boundaries. 
Under the assumption that the GBE for all boundaries is equal (isotropic GBE), the kernel $K_{\delta t}^{i,j}:\R^3 \to \R$ is the same for all grain boundaries and is equal to a Gaussian
\begin{equation}
    \label{eqn:isotropic_kernel}
    K_{\delta t}^{i,j} (\bfx) = \frac{1}{(4 \pi \delta t)^{3/2}} \exp{\left(- \frac{|\bfx|^2}{4 \delta t} \right)}
\end{equation}
which is spherically symmetric and decays with the same rate in all directions. 

Extending this algorithm to the cases where the GBE, $\sigma$, is only a function of misorientation, $\Delta \bfg$ (three parameters model), is straightforward. In this case, all the directions for the grain boundary normal are energetically equally favorable, hence a spherically symmetric kernel for the evolution of each interface can be used. To distinguish the difference between the energy of different grain boundaries due to the misorientation, this spherically symmetric kernel can be scaled according to its energy and defines $K_{\delta t}^{i,j}$ for the boundary with adjacent grains $i, j$ as \cite{esedoḡ2015threshold}:  
\begin{equation}
    \label{eqn:misorientation_kernel}
    K_{\delta t}^{i,j} (\bfx) = \frac{\sigma^{i,j} (\Delta \bfg ^{i,j})}{(4 \pi \delta t)^{3/2}} \exp{\left(- \frac{|\bfx|^2}{4 \delta t} \right)}.
\end{equation}
The Read Shockley GBE \cite{read1950dislocation} is an example of this case.

However, in the case that the GBE also depends on the inclination of the grain boundary, specific grain boundary normals are more favorable than others. Hence, a kernel that can capture this effect should not be spherically symmetric \cite{merriman2000convolution} which brings in more challenges compared to the previous case of the three parameters model. 
Defining this kernel to result in the grain boundary velocity proportional to GBE has been the topic of several studies \cite{bonnetier2012consistency, elsey2018threshold, esedoḡlu2017kernels, ishii1999threshold} and is still an active area of research. 

In this study, we will use the kernel constructed by Bonnetier et al. \cite{bonnetier2012consistency} to simulate the anisotropic evolution of microstructure. 
The Fourier transformation of this kernel $\mathscr{F}\left[K^{i,j}_{\delta t}(\bfx)\right]$ is:
\begin{equation}
    \label{eqn:Bonnetier-kernel}
    \hat K^{i,j}_{\delta t} (\bfxi)= 
    \mathscr{F} \left[K^{i,j}_{\delta t}(\bfx) \right] = 
    \frac{1}{ \delta t^{3/2}}\exp{\left( - \left(\tilde \sigma^{i,j} \left(\delta t \bfxi\right) \right)^2 \right)}, \qquad
    \tilde \sigma(\bfx)^{i,j} =  |\bfx| \sigma^{i,j} \left( \frac{\bfx}{|\bfx|}\right)
\end{equation}
where $\sigma^{i,j}: \mathbb{S}^2 \to \R^+$ is the GBE function for all boundary inclinations for a given misorientation $\Delta \bfg ^{i,j}$ between grain $i,j$, $\tilde \sigma^{i,j}$ is an extension of $\sigma^{i,j}$ such that $\tilde \sigma^{i,j}: \mathbb{\R}^3 \to \R^+$.
Note that for evaluating the convolution step \eqref{eqn:convolution}, in the computational setting we use $f*g = \mathscr{F}^{-1} \left[ \mathscr{F}[f] \mathscr{F}[g]\right]$, so there is no need for computing this kernel in the physical domain \cite{peng2021computational}.
Furthermore, in the kernel used here, the mobility of the interface is embedded such that it is equal to the GBE. There are recent attempts to derive more general kernels where mobility and GBE can be assigned independently, and this is still an active area of the research \cite{esedoḡlu2017kernels}.

According to Algorithm \ref{alg:TD},
using an anisotropic kernel will only affect the convolved value $\psi_i^k$ which is the input to the thresholding step. 
The convolved value $\psi_i^k$ is the result of the convolution between the anisotropic kernel and the characteristic function $\mathbf{1}_{\Sigma_{j}^{k}}$.
Hence, a simplified extension of the Algorithm \ref{alg:TD} to the five-parameter anisotropic GBE can be also achieved by changing the characteristic function $\mathbf{1}_{\Sigma_{j}^{k}}$ according to the GBE and keeping the kernel spherically symmetric Gaussian. Nino and Johnson \cite{nino2023influence} achieved this by replacing $\mathbf{1}_{\Sigma_{j}^{k}}$ with $\sigma^{i,j}(\bfn) \mathbf{1}_{\Sigma_{j}^{k}}$. 
Although methods based on defining an anisotropic kernel following the grain boundary energy and mobility anisotropy are derived from energy minimization \cite{esedoḡlu2017kernels}, more study is required to understand if the simplified version of Nino and Johnson \cite{nino2023influence} is indeed equivalent to a weighted mean curvature flow and minimizes the energy.

\subsection{Grain boundary energy}

Experiments show that the grain boundary energy is a function of five macroscopic parameters, i.e. lattice misorientation between the two adjacent grains (three degrees of freedom), and the inclination of the grain boundary plane (two degrees of freedom) \cite{rohrer2011grain}. 
There are different methods for the representation of these five parameters. 
The most common way is to represent the misorientation between the adjacent grains and plane boundary inclinations separately. 
For example, from experimental measurements considered in this study, for each grain, the rotation of the lattice of each grain relative to a fixed sample frame coordinate system is measured and is represented through a set of Euler angles $(\phi_1, \Phi, \phi_2)$ for rotation around the $(Z,X,Z)$ axes. 
Once the Euler angles are given, the rotation matrix $\bfg_i$ for rotating the sample frame to the frame of grain $i$ can be computed, and $\Delta \bfg = \bfg_i \bfg_j^T$ will give the transformation of the lattice of grain $j$ to the lattice of grain $i$ which is a representation of misorientation between grains $i$ and $j$. 
Independent of misorientation measurements and calculations, the inclination of each point in the grain boundary plane $\bfn$ is computed after the reconstruction and triangulation of grain boundaries in DREAM.3D software \cite{groeber2014dream}. 
Hence, a full five-parameter representation of the grain boundary is given through a normal vector $\bfn$ and a transformation matrix $\Delta \bfg$.

In this study, we use the BRK energy function to evaluate the GBE for any given five parameters for a grain boundary in Ni. 
The BRK GBE function is a non-linear interpolation among 388 different measured grain boundary energies, provided through a MATLAB function as supplementary data in \cite{bulatov2014grain}.
This function is a piece wise interpolation that extends from each cusp in the energy landscape and is consistent with the symmetry of the material. 

Despite the mentioned representation from the experimental data where grain boundary misorientation and plane boundary inclination are represented separately, the input of BRK GBE model is two three-by-three matrices labeled $\bfP$ and $\bfQ$ in which both grain boundary misorientation and plane boundary inclination are combined and represented through these two rotation matrices. A detailed procedure for converting Euler angles and normal vector to the $\bfP \bfQ$ representation is presented in Appendix A.

\subsection{Model validation}
\begin{figure*}[ht!]
	\subfloat[$m=1, \alpha = -0.5$]{\label{fig:validation-1}\includegraphics[width=0.2\textwidth]{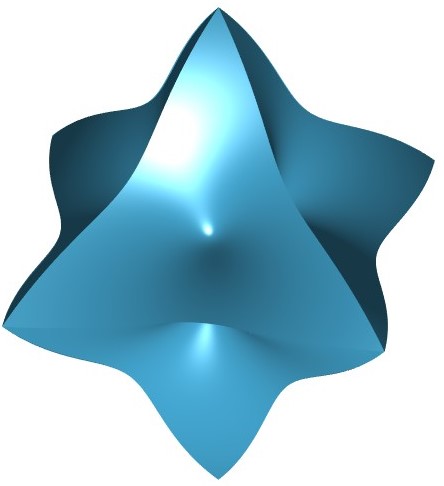}
	 \hfill 
 \label{fig:validation-2}\includegraphics[width=0.2\textwidth]{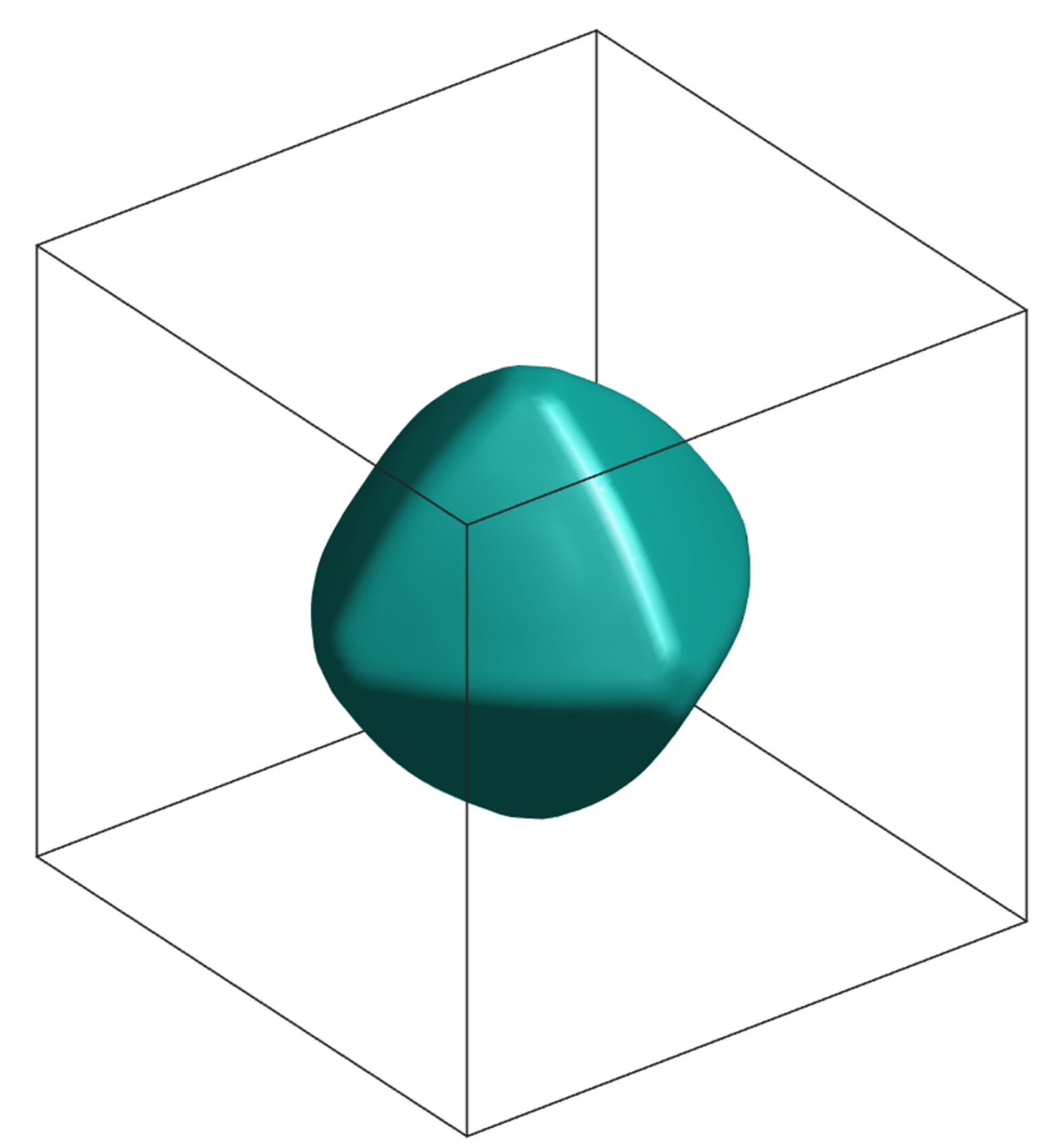}}
    \qquad  \qquad
	\subfloat[$m=4,\alpha=-0.7$]{\label{fig:validation-3}\includegraphics[width=0.2\textwidth]{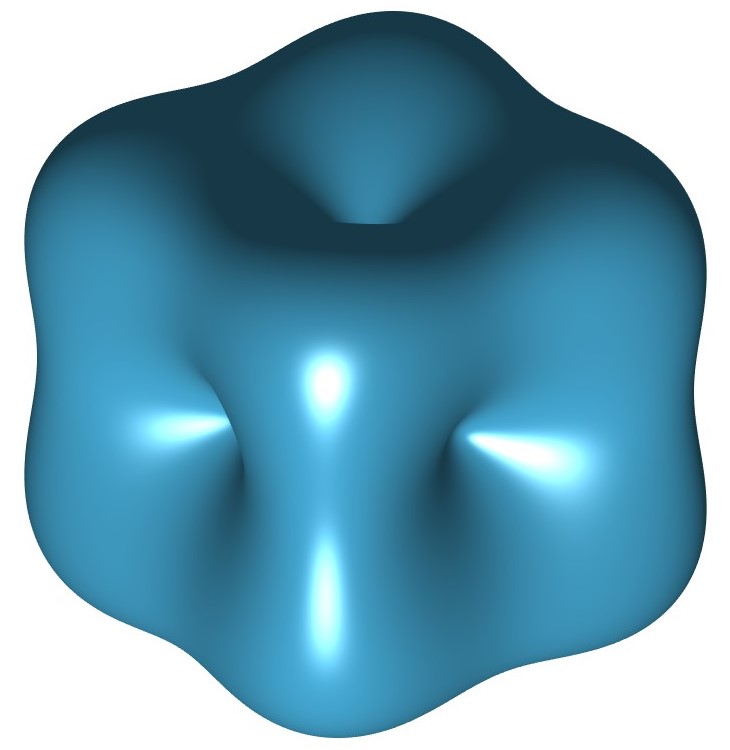}
	 \hfill
 \label{fig:validation-4}\includegraphics[width=0.2\textwidth]{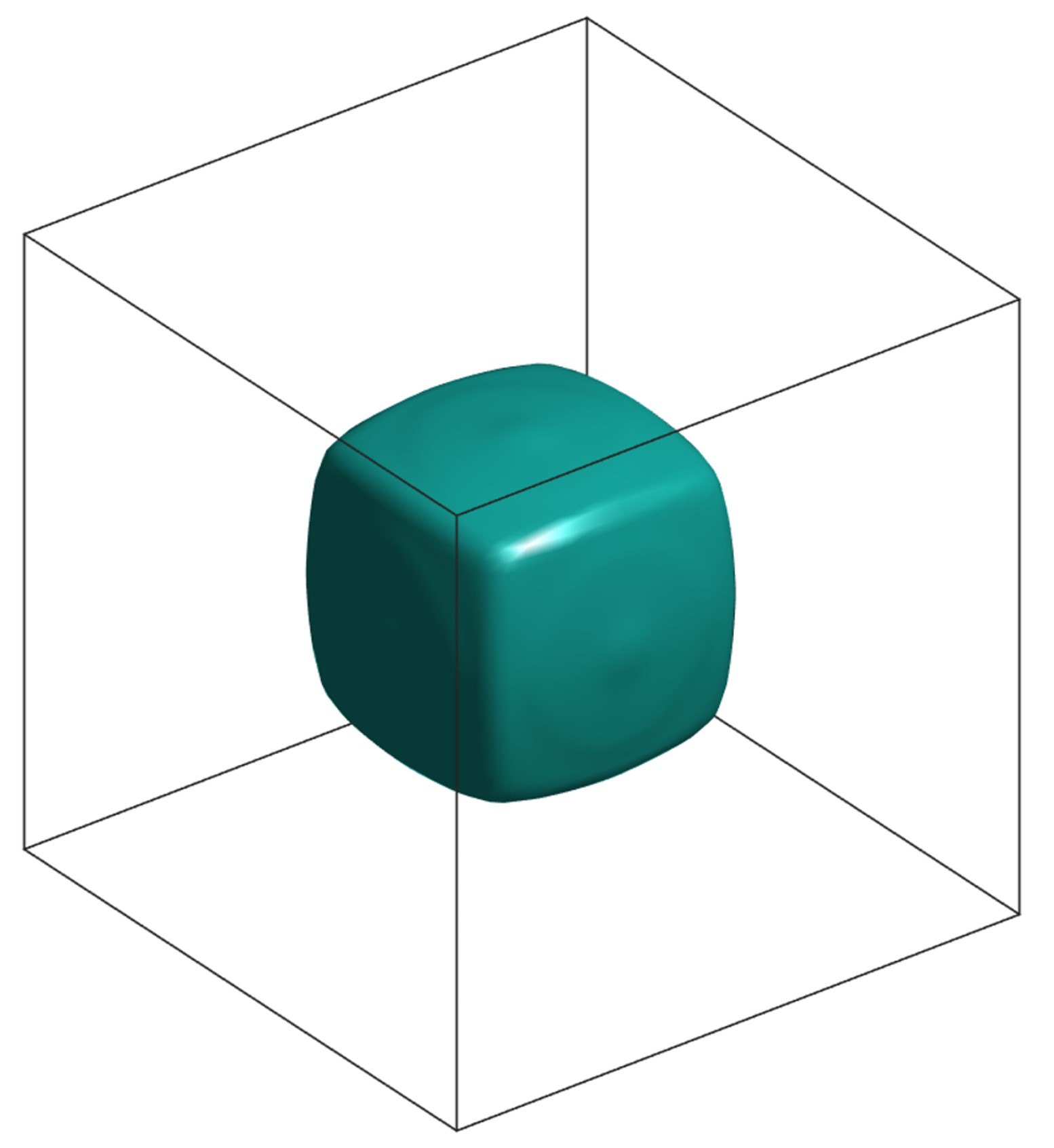}}
	\caption{Simulated equilibrium shape (right figures) of an initially spherical grain in its melt in a box of size $64^3$ (outlined by black lines) using different surface energy (left figures) of the form \eqref{eqn:GBE-validation}.}
	\label{fig:validation}
\end{figure*}

In this section, we validate that the choice of the anisotropic kernel can realistically capture the grain evolution.
We grow a spherical grain located in its melt using different interface energy functions of the form \eqref{eqn:GBE-validation} by using kernel \eqref{eqn:Bonnetier-kernel}. 
The result of our simulation is compared with the result of the front tracking simulations computed by Mohles \cite{mohles20203}. 
Following \cite{mohles20203}, we consider different energy functions of the following form:
\begin{align}
    \label{eqn:GBE-validation}
    GBE = 1 + \alpha \left(|n_x|^m + |n_y|^m + |n_z|^m\right)
\end{align}
where $n_x, n_y, n_z$ are different components of boundary plane normal.
The equilibrium shape of the grain, which is expected to be the Wulff shape of the energy function, is shown in Fig. \ref{fig:validation} and matches the equilibrium shape simulated using the front tracking method (Figure 7 in \cite{mohles20203}) for two sets of parameters $m=1, \alpha = -0.5$, and $m=4,\alpha=-0.7$. 
As it was expected, the inclination of the facets of the equilibrated grains coincides with the direction of minimum energy, ([1 1 1] for the GBE in Figure \ref{fig:validation} (a), and [1 0 0] for the GBE in Figure \ref{fig:validation} (b)).
Hence, the choice of the anisotropic kernel can capture the expected behavior during grain growth.

\begin{figure}[hbt!]
    \centering
    \includegraphics[width=0.7\textwidth]{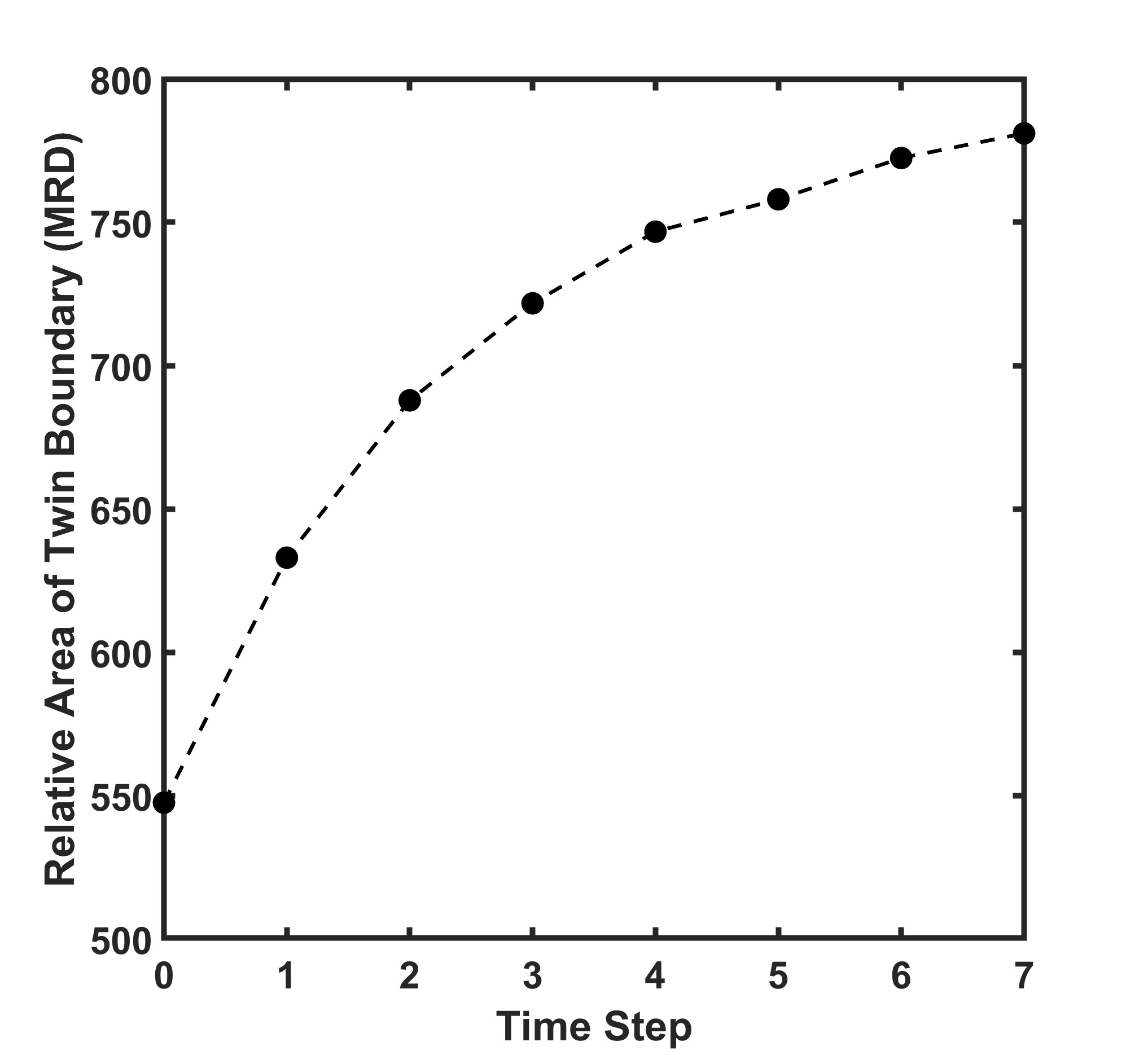}
    \caption{Relative area of twin boundary (MRD) for different timesteps of a simulation started from An4 with the intensity of 547 and the end point of reaching the average grain size of An5.}
    \label{fig:validationMRDAn4}
\end{figure}

Furthermore, a sensitive test of the anisotropic simulation is to observe the evolution of the GB plane distribution, as the GB energy anisotropy influences this \cite{rohrer2023grain}.
Given that the $\Sigma 3$ GB of Ni has the minimum energy at the [111] twin position, an increase in its relative area is expected and observed in the simulation.
Figure \ref{fig:validationMRDAn4} shows the continuous increase in the relative area of twin boundaries for different timesteps of a simulation with the starting point of An4 with the relative area of twin boundaries of 547 multiples of a random distribution (MRD) and the end point of reaching the average grain size of An5, using the BRK energy function (An4 and An5 refer to experimental states defined in section \ref{sec:Result}).

\subsection{Computational challenges of anisotropic simulation}
One main difference between the isotropic and anisotropic simulation is that in the isotropic simulation the kernel is the same for all grain pairs, and hence it can be computed once at the start of the simulation and used at any time later during the simulation.
However, in the full anisotropic simulation, the convolution kernel is different for each grain pair and there is no linear relation between kernels, as in the case of GBE only being a function of misorientation.
Hence, a key challenge is that the experimental volume contains a large number of different GB types with an approximate number of distinct grain boundaries of 34000 in each microstructure. 
Additionally, to evaluate the non-spherically symmetric anisotropic kernel \eqref{eqn:Bonnetier-kernel} for each two-grain pair with a given grain boundary misorientation, the GBE for all inclinations of the boundary plane is required. 

While the energies are potentially available from the BRK function, evaluating them for all points in a kernel is prohibitively expensive numerically. 
Hence, we define a coarser grid including 6192 different normal vectors distributed uniformly on a sphere. For each grain boundary misorientation, we store the energy at these 6192 different inclinations, at the beginning of the simulation. We use the nearest interpolation method to compute energy values on the finer grid of the simulation.

\subsection{Experimental data and simulations}
This paper aims to compare the experimentally observed microstructure evolution of a high-purity Ni sample during annealing with simulated microstructures using both isotropic and anisotropic grain boundary energy. The sample was measured at six different times using near-field high energy X-ray diffraction microscopy \cite{hefferan2010statistics,hefferan2012measurement}. The sample underwent annealing for about 30 minutes at $800^{\circ}$C between each measurement. Previous publications have outlined the specifics of data acquisition and interpretation \cite{hefferan2010statistics,li2011imaging,hefferan2010tests}.
Six repeated measurements of the same sample volume were used to reconstruct the shapes and orientations of grains after successive annealing treatments
\cite{li2011imaging,hefferan2010tests}. The data are represented as a set of discrete voxels using DREAM.3D \cite{groeber2014dream, bhattacharya2019three}. We refer to them as An0, An1, An2, An3, An4, and An5 throughout the paper. 
The microstructures contained 2400 to 3000 grains
made up of voxels with dimensions of 2.3 × 2.3 × 4.0 $\mu $m$^3$. 
In the initial state, there was an average of 2347 voxels per grain.
We note that the data used here is the same that was used in our previous isotropic simulation \cite{peng2022comparison}, with one important difference. In the previous work, all twinned grains were merged to form single grains, in an effort to ameliorate the known anisotropy of the energy. In contrast, all twins were preserved in the present work.

The first five reconstructed microstructures (An0 to An4) were considered as an initial state of different simulations independently. Both isotropic and anisotropic simulation was performed for each of the five initial states. 
The average grain size increases throughout the simulation and experimental annealing, and the simulations were terminated when the average grain size reaches the average grain size in the next experimental anneal step \cite{peng2022comparison}. 
All grain boundary properties (relative area, curvature, velocity) were calculated using methods described in previous publications \cite{bhattacharya2021grain, xu2024grain}.

\subsection{Grain boundary properties}
The analysis of the experimental data and simulations involved calculation of the grain boundary relative areas, curvatures, and velocities.  The calculations begin with converting the voxelated grain boundaries to meshed interfaces in DREAM.3D \cite{groeber2014dream}.  
The relative areas of particular boundaries were determined using the method of Glowinski and Morawiec \cite{glowinski2014analysis}.  
The grain boundary curvature was the area weighted mean curvature of all of the mesh elements belonging to a certain boundary, computed by DREAM.3D \cite{zhong2017five,xu2024grain}.  
The migration velocity of each boundary was calculated based on the volume of voxels exchanged across the boundary; detailed information can be found in the previous publications \cite{xu2024grain, bhattacharya2021grain}.


\section{Results}
\label{sec:Result}
\begin{figure*}[ht!]
	\subfloat[Initial state (An0)]{\label{fig:Microstructure-1}
        \includegraphics[width=0.22\textwidth]{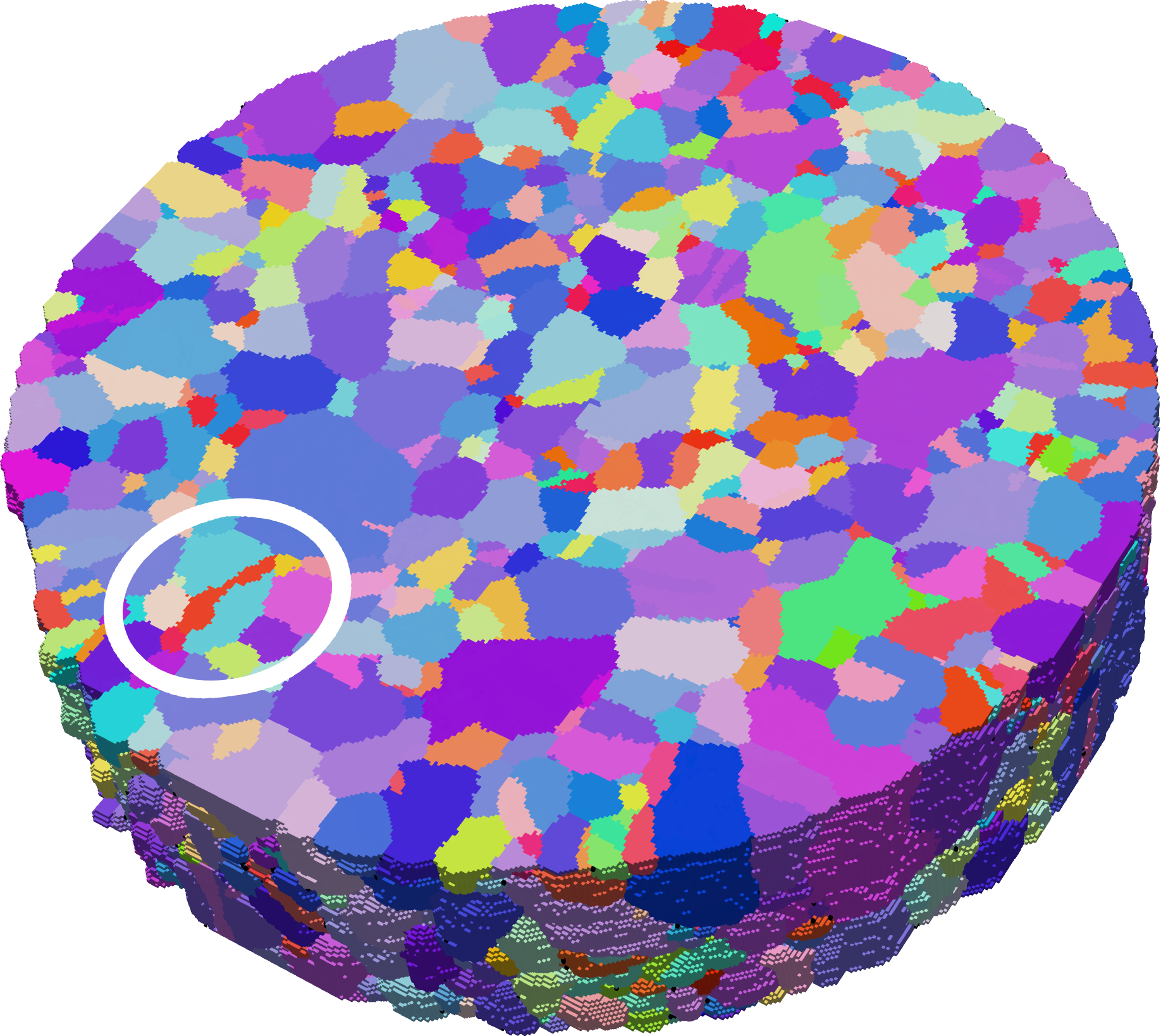}}
	\hfill
	\subfloat[Experiment (An1)]{\label{fig:Microstructure-2}\includegraphics[width=0.22\textwidth]{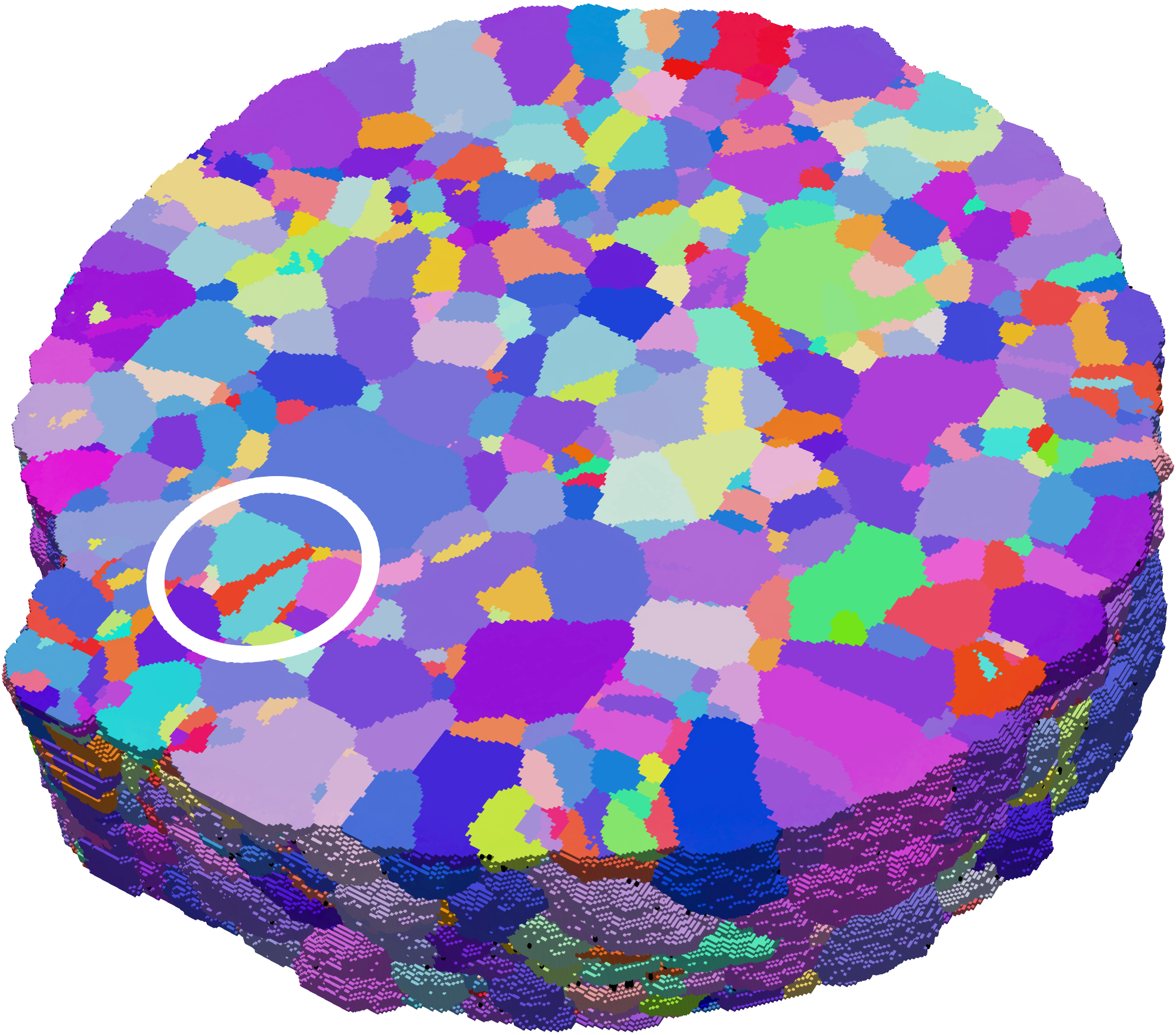}}
        \hfill
	\subfloat[Isotropic simulation]{\label{fig:Microstructure-3}\includegraphics[width=0.22\textwidth]{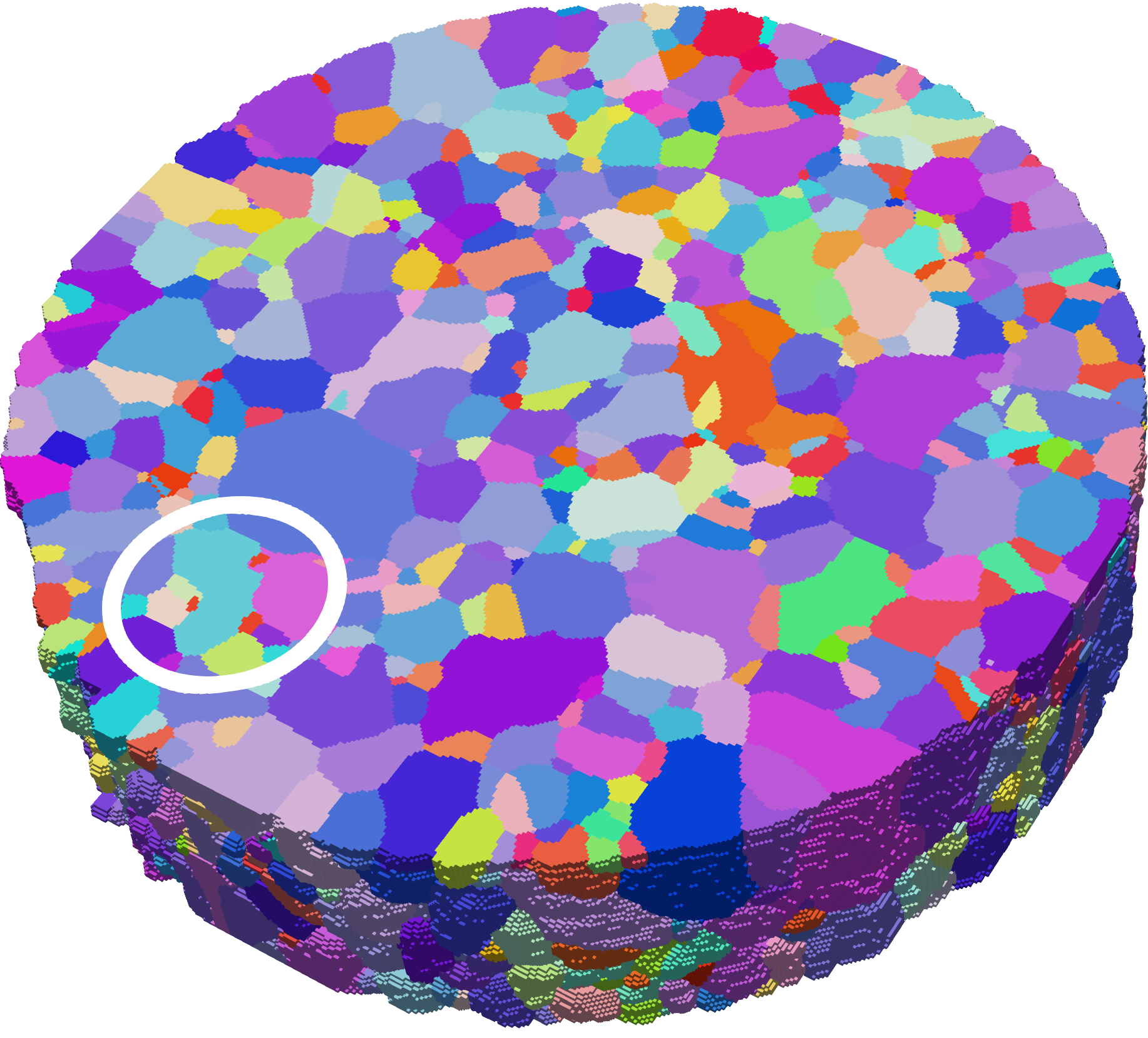}}
	\hfill
	\subfloat[Anisotropic simulation]{\label{fig:Microstructure-4}\includegraphics[width=0.22\textwidth]{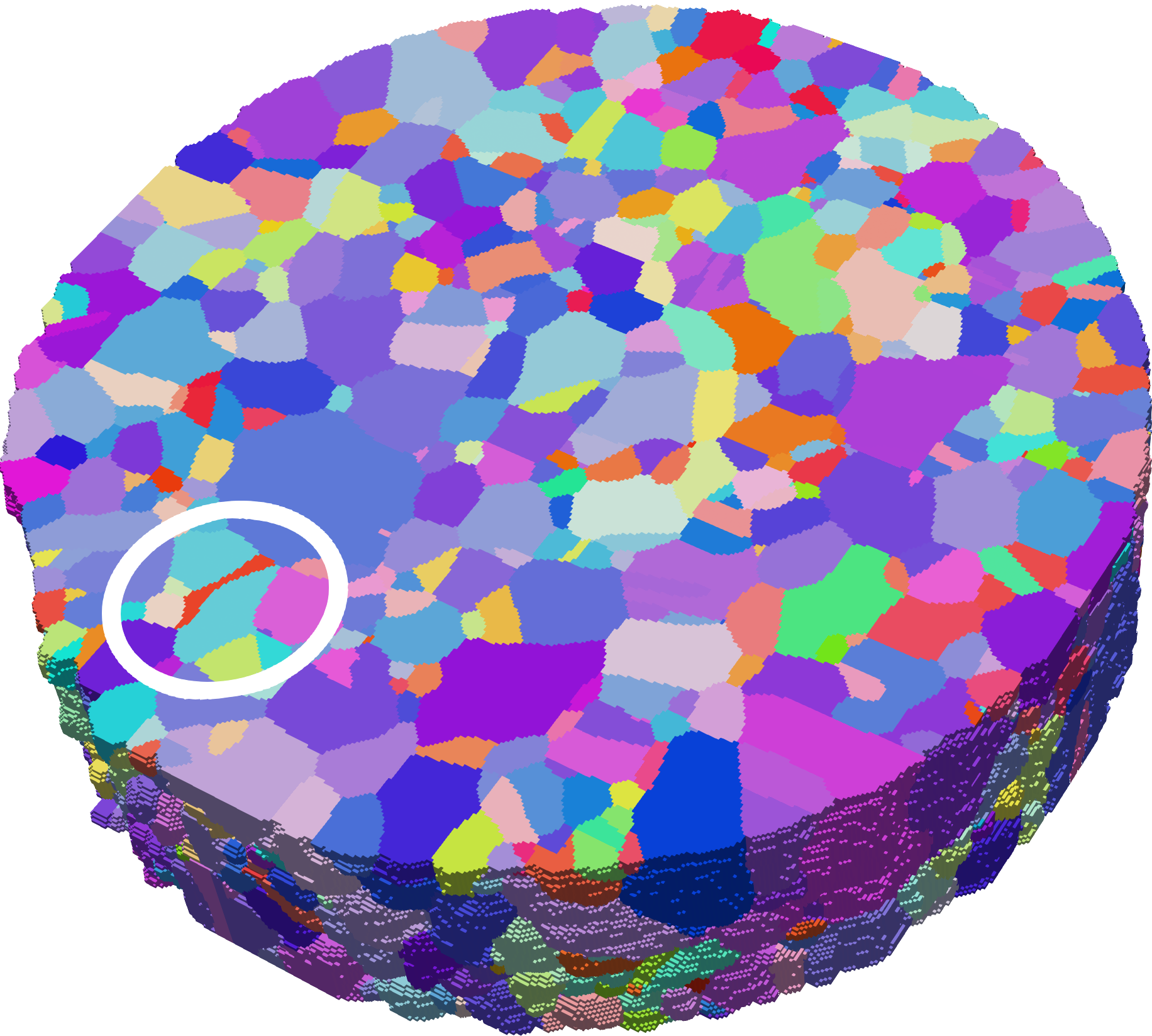}}
	\caption{Experimentally measured and simulated microstructures. In each case, the diameter of the cylinder is about 1 mm.}
	\label{fig:Microstructure}
\end{figure*}

The cylindrical Ni sample in the initial experimental state (An0), and An1 are depicted in
Figure \ref{fig:Microstructure} (a) and (b), where the 2972 and 2669 grains are colored by orientation.
Figure \ref{fig:Microstructure} (c) and (d) show the evolved microstructure from An0 to An1 using isotropic and anisotropic simulations, respectively. 
A cursory comparison shows only small differences between the four microstructures. 
However, one
exemplary feature is highlighted by the white oval. A twin (red) bisects a blue colored grain. 
In the experiment and in the anisotropic simulation, the twin is preserved. 
However, in the isotropic simulation, it is eliminated. 
This is because when all grain boundary energies are the same, spheroidal, energy minimizing grain shapes are preferred over grains with high aspect ratios. 
In the remainder of this section, we use distributions of properties to compare the microstructure more systematically.

The behavior of the evolved microstructure during the experiment, isotropic, and anisotropic simulations from different perspectives is compared.  
Since the simulations were performed independently for different anneal stages as the initial state of the simulation, the result of each simulation is only compared with the next experimental anneal step. 
The following notation is considered to present the results in this section:
\begin{itemize}
    \item An0-1: The experiment/simulation started with the An0 microstructure and terminated when the average grain size was equal to the average grain size of the An1 experiment.
    \item Initial state: Experimental data of An0.
    \item Experiment: Microstructure evolved experimentally and stopped at An1.
    \item Anisotropic simulation: The output of the simulation using an anisotropic kernel in Algorithm \ref{alg:TD}. The input is An0 experimental data and the simulation was stopped when the average grain size reached the average grain size of An1.
    \item Isotropic simulation: The output of the simulation using an isotropic kernel in Algorithm \ref{alg:TD}. The input is An0 experimental data and the simulation was stopped when the average grain size reached the average grain size of An1.
\end{itemize}
A similar notation is used for An1-2, An2-3, An3-4, and An4-5. 

\begin{figure}[ht!]
    \centering
    \includegraphics[width=1\textwidth]{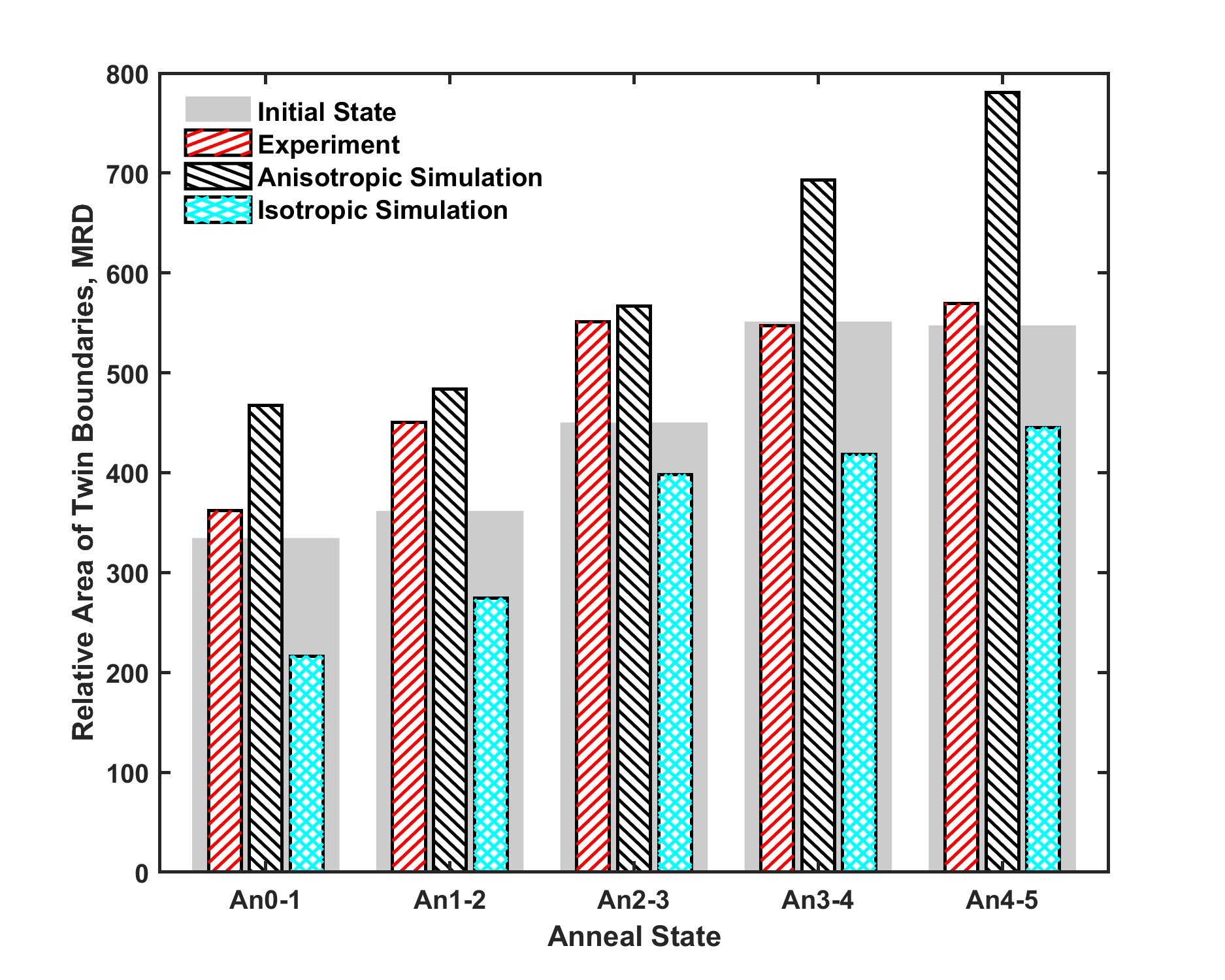}
    \caption{The twin boundary relative area for experimental and simulated data at different anneal states.}
    \label{fig:MRD}
\end{figure}

\begin{figure}[ht!]
    \centering
    \includegraphics[width=1\textwidth]{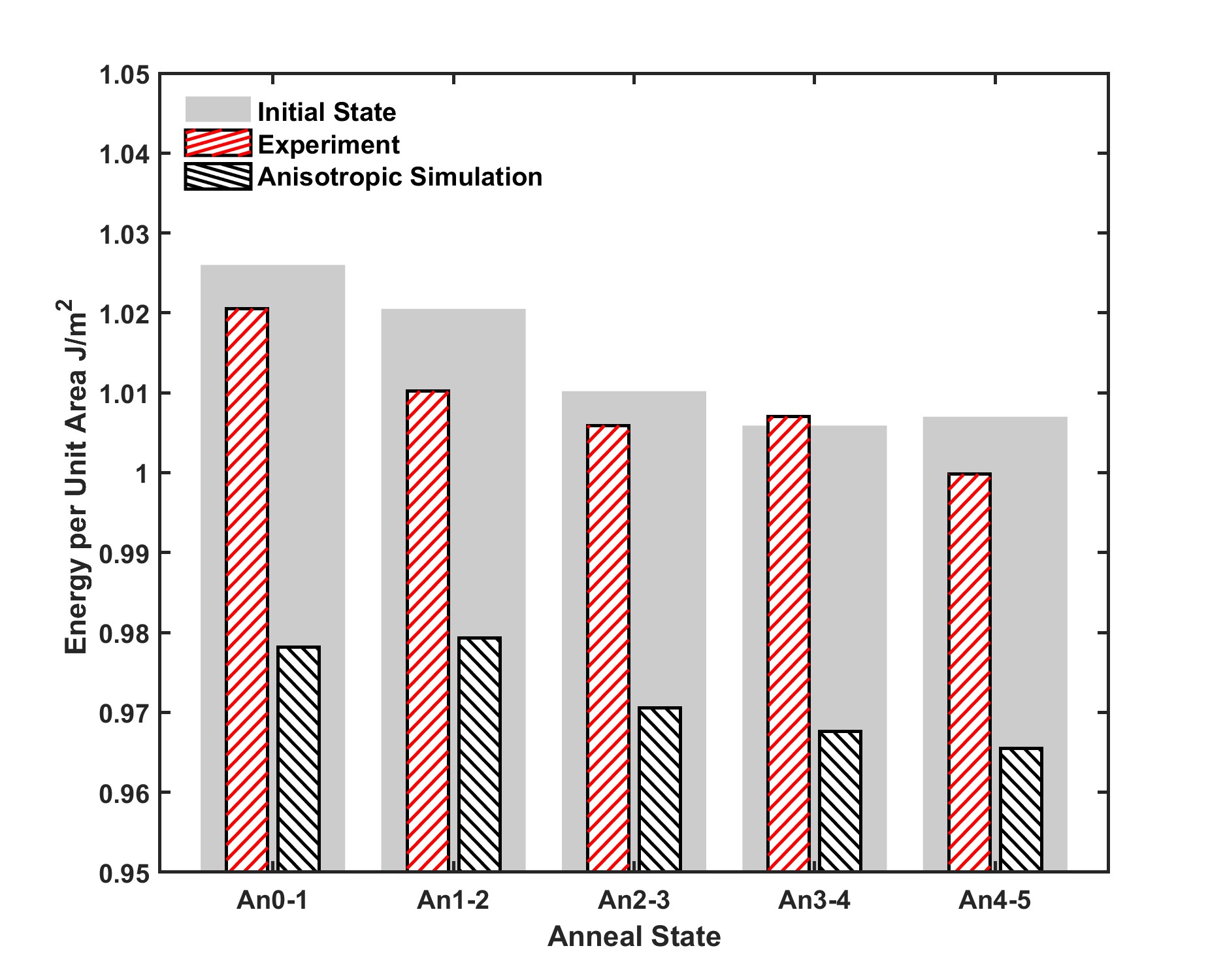}
    \caption{Energy per unit area for experimental and simulated data at different anneal states.}
    \label{fig:EPA}
\end{figure}

Two main statistical features that are expected to be captured in the anisotropic simulation are the energy distribution of grain boundaries and the relative area of $\Sigma 3$ twin boundaries. 
Figures \ref{fig:MRD} and \ref{fig:EPA} 
show the relative area of the twin boundaries for simulated and experimental anneal steps and the microstructure energy per unit area of the grain boundaries.
For each step, compared to the initial state, the relative area of the twin boundaries increases in both the experiment and anisotropic simulations, while it always decreases in the isotropic simulation. 
Similarly, compared to the initial state, the energy per unit area decreases (except for An3-4) and the anisotropic simulation always decreases the energy. 
The small increase in energy for An3-4 might be the result of uncertainties in the experiment and reconstruction.
Note that since the uniform grain boundary energy of one is assigned for all the boundaries in isotropic simulation, the energy per unit area always remains one and is not relevant for the comparison. 
The increase in the relative area of the twin boundary and the decrease in the average grain boundary energy in the anisotropic simulation is always greater than in the experiment; this will be discussed in section \ref{sec:Discussion}.

\begin{figure}[ht!]
    \centering
    \includegraphics[width=1\textwidth]{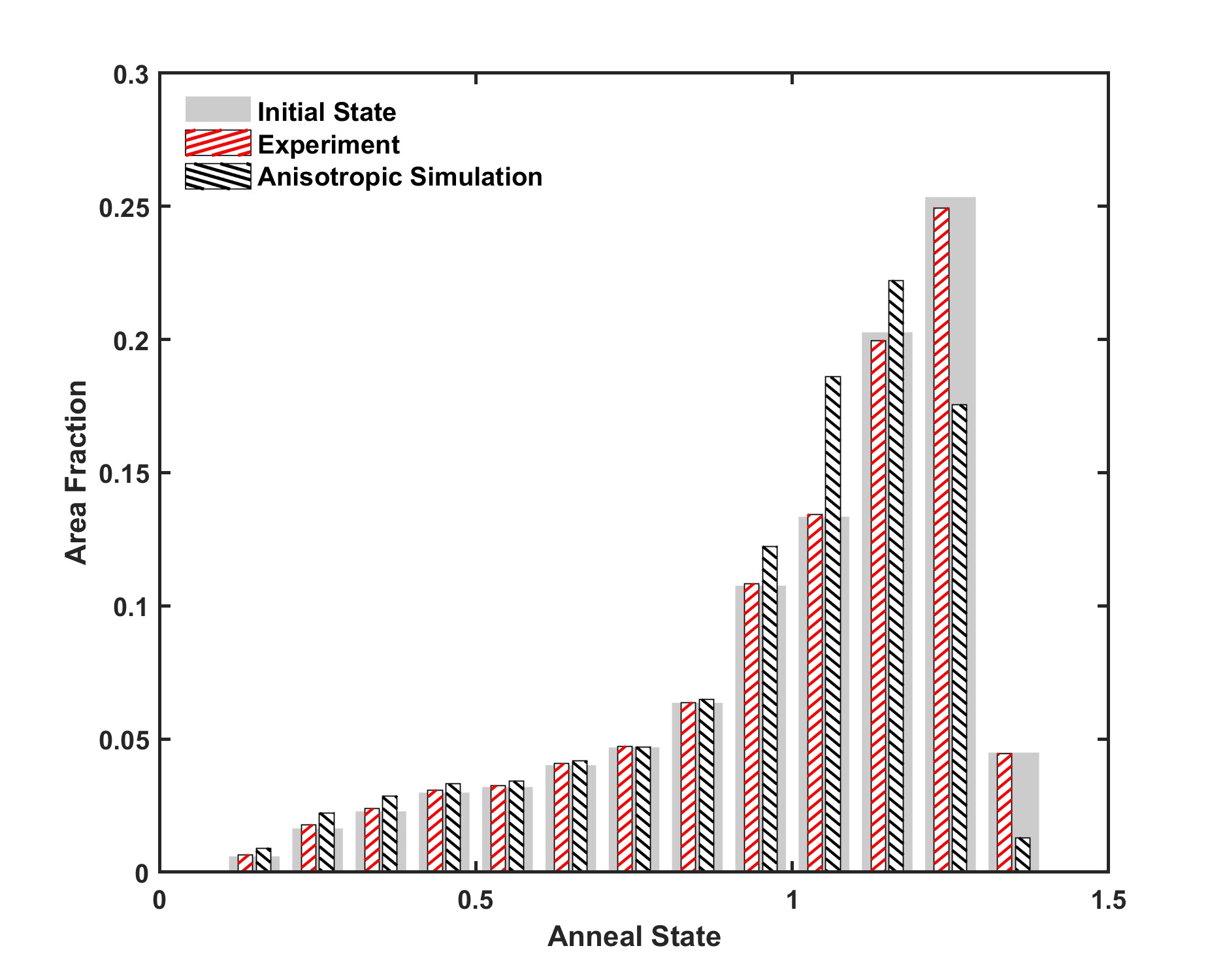}
    \caption{Distribution of simulated and experimental GBEs combined for all anneal states.}
    \label{fig:GBE-distribution}
\end{figure}

Figure \ref{fig:GBE-distribution} looks more closely at the energy distribution of the simulated and experimental data for all anneal stages combined together. 
The thick dark blue bars behind the thinner bars show the energy distribution for the initial state of the experiment and simulation, and the two thinner bars show the GBE distribution for the anisotropic and next experimental anneal state. 
For the lower energy grain boundaries, there are more boundaries in the final states of the experiment and simulation and for the higher energy boundaries, there are
fewer. 
A comparison of these distributions shows that the experiment and anisotropic simulation shift the distributions so that there are more boundaries with lower energy.

\begin{figure}[h!]
    \centering
    \includegraphics[width=0.7\textwidth]{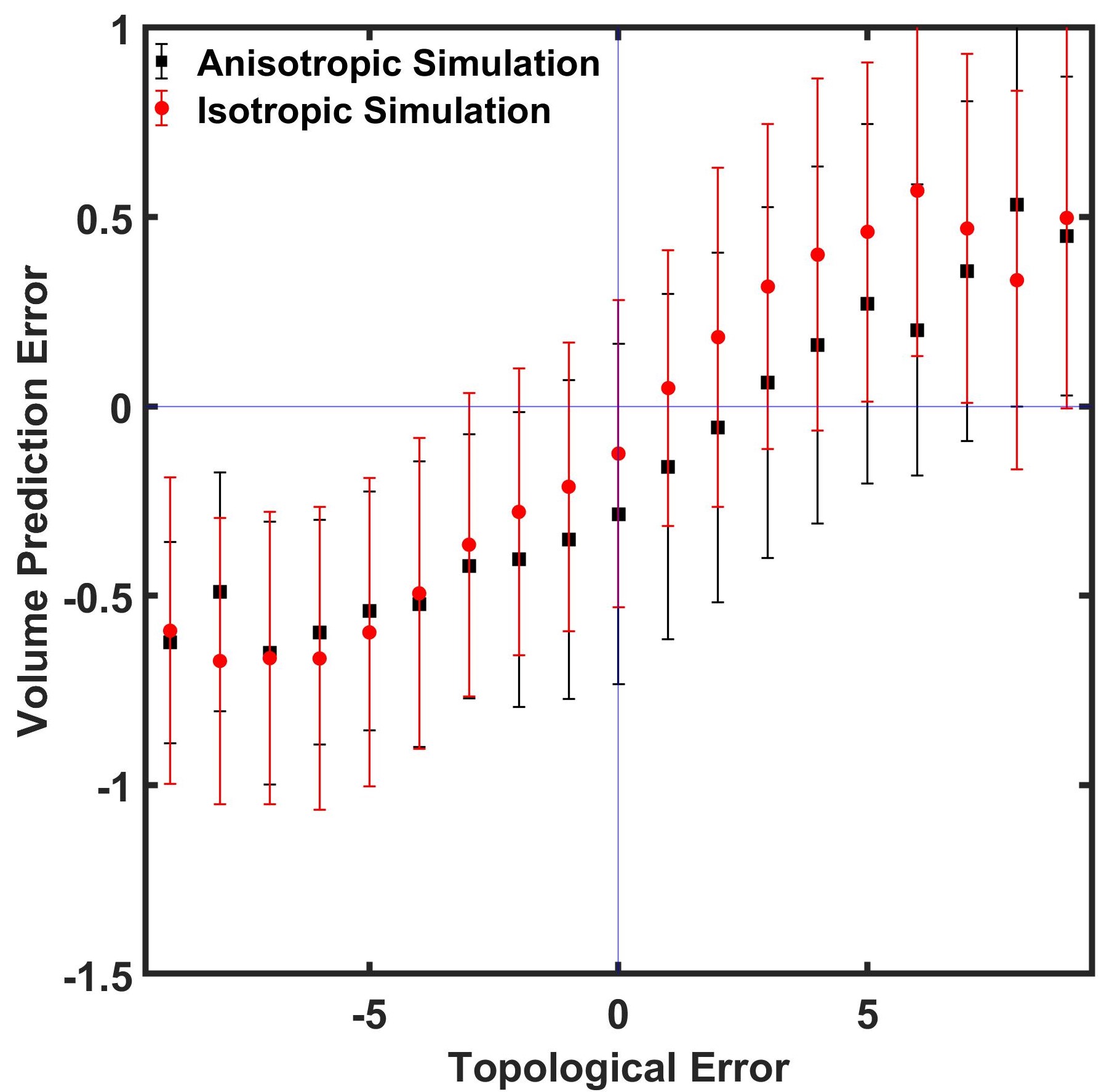}
    \caption{\label{fig:ErrorTopoVol} The volume prediction error ($VPE$) as a function of topological error ($TE$). $VPE$ is the fractional difference in predicted and observed grain volume. $TE$ is the difference in grain face evolution between simulation and experiment.}
\end{figure}

Two sources of differences between the simulated and observed microstructures are differences in the volume changes of grains and differences in the neighborhoods. 
In \cite{peng2022comparison}, it was shown that these effects are correlated.
To examine whether or not this occurs in the current simulation,
we compare the volume prediction error ($VPE$) with the topological error ($TE$) for individual grains. $VPE$ and $TE$ are defined as follows:
\begin{align}
    \Delta N_s &= N_\text{sim} - N_\text{exp(initial)} \\
    \Delta N_e &= N_\text{exp(final)} - N_\text{exp(initial)} \\
    TE &= \Delta N_s - \Delta N_e \\
    VPE &= \frac{V_s - V_e}{V_e}
\end{align}
where $N_\text{exp(initial)}$ is the number of neighbors of grain in the initial experiment state, $N_\text{exp(final)}$ is the number of neighbors of the same grain in the final experiment state, and $N_\text{sim}$ is the number of neighbors of the same grain in the final simulation state. $VPE$ is the fractional difference in volume predicted by the simulation of the final anneal state ($V_s$) and experimental final state ($V_e$). $TE$ is the difference in $\Delta N$ for each grain between simulation and experiment. In other words, $TE$ is the error in predicting topological evolution by the simulation. Figure \ref{fig:ErrorTopoVol} plots the volume prediction error as a function of topological error for isotropic and anisotropic simulations. 
A low $VPE$ indicates a small difference between the final volume predicted and the actual final volume of the grain. A high $TE$ value means there is a large error in predicting the topological evolution of the grains. 
Similar behavior of isotropic and anisotropic simulation in $VPE$ vs $TE$ suggests that considering energy anisotropy does not improve this aspect of the simulation. 

\begin{figure}[h!]
    \centering
    \includegraphics[width=0.7\textwidth]{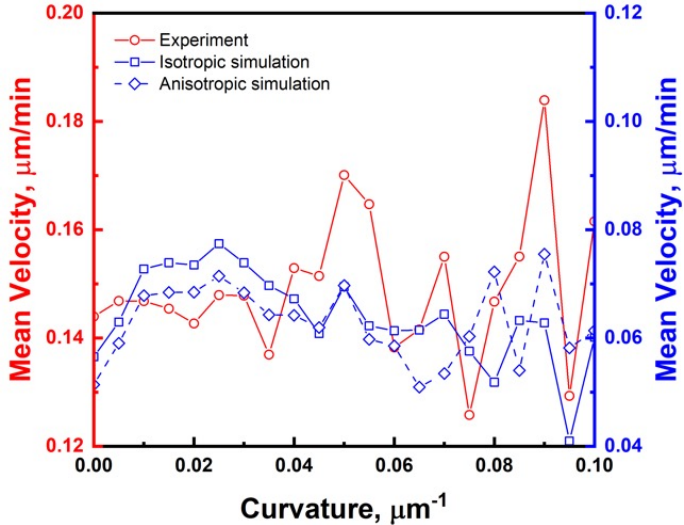}
    \caption{\label{fig:VelocityCurvature} Mean velocity as a function of curvature in experiment and simulations.}
\end{figure}

Measurements of the grain boundary velocity and curvature from the experimental data showed no correlation between these quantities \cite{bhattacharya2021grain}.  
This unexpected result was also reported for $\alpha$-Fe \cite{xu2024grain} and SrTiO$_3$ \cite{muralikrishnan2023observations}.  
As illustrated in Figure \ref{fig:VelocityCurvature}, the simulated data also lacks a correlation between curvature and velocity, consistent with the experiment.  
When interpreting this result, it is important to note that the curvature data is always from the initial (experimental) data, so is independent of the simulation.  
Similarly, the velocity calculation also incorporates information from the initial state, so this undoubtedly influences the result.  
It is somewhat surprising the isotropic simulation does not show a correlation between velocity and curvature.  When the same method was used to simulate the evolution of $\alpha$-Fe with isotropic grain boundary energies, a strong correlation between velocity and curvature was found \cite{xu2024grain}.  The one difference is that the grain shapes in the $\alpha$-Fe were equiaxed, while the Ni microstructure contains many non-equiaxed shapes that result from twinning.  
Instantiating the simulation with this structure "imprints" this anisotropy in the microstructure and this is apparently enough to disrupt any correlation between velocity and curvature that the simulation might otherwise produce.  Therefore, the absence of a correlation between curvature and velocity in the anisotropic simulation can not solely be attributed to the grain boundary energy anisotropy.

\section{Discussion}
\label{sec:Discussion}
The results of the simulations reproduce the decrease in the average energy of grain boundaries through grain boundary replacement \cite{xu2023energy}, a key phenomenon found in the experiment.  
The simulations show that the decrease in the average energy is associated with a decrease in the fraction of high energy grain boundaries and an increase in the fraction of low energy grain boundaries.  
A key finding of this work is that this phenomenon emerged simply by introducing an anisotropic energy distribution.  
In other words, there was no need to introduce a new physical mechanism in the model.  
The simulation is constructed to reduce the total area.  
The results suggest that when multiple possible grain boundary migration paths are possible, on average, the one that is selected is the one that reduces the area in such a way that lower energy grain boundaries are increased in area at the expense of higher energy boundaries.  This process leads to the results in Figures \ref{fig:EPA} and \ref{fig:GBE-distribution}.
The accumulation of low energy grain boundaries during grain growth has been observed in experiment \cite{rohrer2011grain} and in simulations \cite{gruber2005effect, kim2014phase, nino2023influence} before, but this is the first direct comparison of experiment and simulation showing that observed features of the microstructure emerge by assuming realistic anisotropic energies.

One important difference between the experiment and the simulation is that the grain boundary replacement process is more significant in the simulation.  For example, the decrease in the average energy at each time step in the experiment is < 1\% while in the simulation it was on the order of 4\% (See Figure \ref{fig:EPA}).  
The most likely source of this difference is the difference between the energy anisotropy in the simulation and experiment.  
The simulation used the BRK energies at 0 K \cite{bulatov2014grain}, while the experiment \cite{hefferan2010statistics} was carried out at 1073 K.
The differences in the energies among boundaries is certainly smaller at 1073 K, and this is expected to decrease the driving force for the grain boundary replacement process.  Simulations conducted with scaled energies showed that decreasing the energy differences slowed the decrease in the average energy, with the isotropic case presented here being the extreme example.  However, scaling the energies in a realistic way proved challenging.  
The assignment of temperature dependent grain boundary properties would make it possible to simulate the effect of temperature on microstructure evolution.

When comparing the results of the simulation and experiment, one should keep in mind that the simulation is instantiated with experimental data, seeding the process with the ground truth at that time step.  
Such a simulation is obviously better situated to reproduce the experiment than starting from a non-physical state.  While one might guess that this guarantees the simulation produces a realistic microstructure, the results show otherwise.  
As illustrated in Figure \ref{fig:MRD}, the simulation with isotropic grain boundary energies evolves the microstructure in the wrong direction; the relative areas of twin boundaries decreases with time while the experiment and anisotropic simulation both increase the relative areas of the twins.  
In other words, even though the simulation is provided with the correct starting point, it evolves in the wrong direction. 
The option of instantiating the simulation with a starting point different from the observed microstructure could also be informative, but this seems less likely to lead to a better understanding of the physical process that occurs in real grain growth.

Grain-by-grain comparisons of microstructure evolution have been unsuccessful in the past \cite{mckenna2014grain, bhattacharya2019three, zhang2018three} and the implementation of anisotropic energies has not improved the situation, as illustrated in Figure \ref{fig:ErrorTopoVol}.  
The basic problem is that as soon as a single critical event (the disappearance of a grain face for example) is predicted incorrectly, the microstructure evolves along a different path.  
The energy distribution used in the simulation is thought to be a reasonable approximation of the energies at 0 K, but this approximation deviates from the true energy distribution at the experimental temperature, and this might contribute to differences in the evolutions.  
Even if the energy was completely accurate, there is evidence that some aspects of grain boundary migration are not entirely reproducible in atomistic simulations \cite{qiu2023variability}. 
In other words, when grain boundary migration is simulated many times by molecular dynamics, the outcome is not fully reproducible.  
If so, there is no possibility of reproducing the exact sequence of critical events in microstructure evolution, even if the physical process in the experiment is fully deterministic.    

The observed reduction in grain boundary energy provides an additional energy dissipation mechanism during grain growth, as described previously \cite{xu2023energy}.
This is an additional driving force that influences grain boundary migration and is absent in simulations with isotropic grain boundary energies.  
Previous reports that the grain boundary character distribution evolves in response to assumed anisotropic energies \cite{gruber2005effect, kim2014phase, salama2020role, nino2023influence} and the results presented here that the assumption of realistic energies leads to simulated results that reproduce many features of the experiment indicate that anisotropic grain boundary energies are required input for realistic simulations.  While this seems to add a complexity to the simulations, realistic, five-parameter, grain boundary energy functions for the fcc \cite{bulatov2014grain} and bcc  \cite{chirayutthanasak2024universal} structures are available and, at least for the TD simulation, it is not necessary to alter the energy minimizing procedure.  

\section{Conclusion}
We have compared the experimentally observed microstructure evolution of a Ni sample with isotropic and anisotropic simulations. 
In the anisotropic simulation, the grain boundary energies
were defined by the BRK energy function.
The assumption of anisotropic grain boundary energies leads to an increase in the relative areas of low energy twin boundaries and a
change in the grain boundary energy distribution that reduces the average grain boundary energy. 
These changes result from the
anisotropic grain boundary energy, without any changes in the energy minimizing algorithm, and do not occur when isotropic
energies are assumed. 
The results indicate that realistic simulations of grain growth in polycrystals require anisotropic grain
boundary energies that approximate those in the real material.

\section*{Software and Data Availability}

A version of the code developed for this work is available at \url{https://github.com/Kiana-Naghibzadeh/TD_aniso_BRK}. 
And, the data used for this work are available at 
\url{http://mimp.materials.cmu.edu/~gr20/Grain_Boundary_Data_Archive/Ni_velocity/Ni_velocity.html}

\begin{acknowledgments}
    This work was supported by the National Science Foundation under DMREF Grant No. 2118945. 
    We acknowledge NSF for XSEDE computing resources provided by Pittsburgh Supercomputing Center.
    Kiana Naghibzadeh was partially supported by the Bushnell Fellowship.
    This research used resources of the Advanced Photon Source, a U.S. Department of Energy (DOE) Office of Science User Facility operated for the DOE Office of Science by Argonne National Laboratory under Contract No. DE-AC02-06CH11357.
\end{acknowledgments}

\makeatletter
\renewcommand*{\thesection}{\Alph{section}}
\renewcommand*{\thesubsection}{\thesection.\arabic{subsection}}
\renewcommand*{\p@subsection}{}
\renewcommand*{\thesubsubsection}{\thesubsection.\arabic{subsubsection}}
\renewcommand*{\p@subsubsection}{}
\makeatother

\appendix
\section{Euler angles and normal vector to  \texorpdfstring{$\bfP \bfQ$}{PQ} representation}
Matrices $\bfP$ and $\bfQ$ are rotation matrices from the frame of grain $i$ and $j$ to a reference frame where the normal of the boundary plane is aligned with the x-axis, respectively. Hence, 
\begin{enumerate}
    \item The first rows of $\bfP$ and $\bfQ$ represent the normal boundary plane in the frame aligned with the lattice of grain $i$, and the lattice of grain $j$, respectively.
    \item The rotation matrix from the lattice of grain $j$ to the lattice of grain $i$, i.e. $\Delta \bfg = \bfg_i \bfg_j^T$, is equal to $\bfP^T \bfQ$.
\end{enumerate}

Given a boundary plane normal $\bfn$ represented in the sample frame from triangulation, the first row of $\bfP$ is $\bfg_1 \bfn$. Since $\bfP$ is a rotation matrix, all its rows should be perpendicular to each other, hence the second row of $\bfP$ can be any normalized vector perpendicular to the first row. 
The third row is perpendicular to rows 1 and 2, i.e., the cross product of row 1 and row 2. Finally, $\bfQ$ can be computed using the equality of $\Delta \bfg = \bfP^T \bfQ$.
Note that the second row of matrix $\bfP$ is not unique, hence $\bfP$ and $\bfQ$ are not unique, but any $\bfP$ and $\bfQ$ that satisfies conditions 1 and 2 will result in the same energy value.

\newcommand{\etalchar}[1]{$^{#1}$}

\end{document}